# The physics of randomness and regularities for languages (lifetimes, family trees, and the second languages); in terms of random matrices

Çağlar Tuncay
Department of Physics, Middle East Technical University
06531 Ankara, Turkey
caglart@metu.edu.tr

**Abstract:** The physics of randomness and regularities for languages (mother tongues) and their lifetimes and family trees and for the second languages are studied in terms of two opposite processes; random multiplicative noise [1], and fragmentation [2], where the original model is given in the matrix format. We start with a random initial world, and come out with the regularities, which mimic various empirical data [3] for the present languages.

**1. Introduction:** In a recent work [4] the cities and the languages are treated differently (and as connected; languages split since cities split, etc.); thus two distributions are obtained in the same computation at the same time. In the present contribution we consider only the languages, where our essential aim is to show that the several regularities about the languages may be evolving out of randomness; and, the present processes may be shaping the evolution. Following section is the original model in matrix format; the next section is for application and Section 4 is for the results. Discussions and conclusion is the last section.

**2. Model:**

*2.1 Initiation:* At the beginning of our time ($t$), we have $m(t=0)$ languages, where the number of speakers are ($L(t=0)_i$, $i \leq m(0)$) defined randomly; and, each of the ancestor language belongs to an ancestor family $f(t=0)$ ($\leq m(0)$). Adults choose a language as a second language if it is bigger than the mother language in size.

*2.2 Evolution:* At each time step $t$, the languages split (fragment) with the probability $h$, where the splitting ratio is $s$. The fragmentation (splitting, mutation) means that, if the current population of a language ($i$) is $L(t)_i$, $sL(t)_i$ many members form another population and $(1-s)L(t)_i$ many continue to speak the same language. It is obvious that, the results do not change if the mutated and surviving members are interchanged, i.e., if $1-s$ is substituted for $s$. The number of the languages $m(t)$ increases by one if one language splits; if any two of them split at $t$, then $m(t)$ increases by two, etc.

The languages ($L(t+1)$) at a time $t+1$ evolve out of these ($L(t)$) at $t$;

$$L(t+1) = E(t)L(t)  , \qquad (1)$$

where $L(t)$ is a ($1 \times m(t)$) matrix, and $E(t)$ is a rectangular ($m(t+1) \times m(t)$) matrix, which represents the evolution of the languages in a time step from $t$ to $t+1$ (evolutor). The number of the current languages $m(t)$ increases randomly, where one may have

$$m(t+1) - m(t) \propto r_m\, h\, m(t)  , \qquad (2)$$

for a random number $r_m$ which is uniformly distributed between zero and one, ($0 \leq r_m < 1$) and for a given constant $h$, which is small. It is obvious that, $m(t)$ increases exponentially (up to randomness) with time, where the exponent is proportional to $h$. We may have a similar relation for the families $f(t)$ ($f(t) \to m(t)$), in terms of $h_{fam}$ ($h_{fam} \to h$); where, $h_{fam}$ is the fragmentation rate for the families, and the number of the families ($f(t)$) increases exponentially (with the exponent, which is proportional to $h_{fam}$) in time.

In Eq. (1) we define $E(t)$ as:

$$E(t)_{ji} = \delta_{ji} + r\, r_{ji} \quad , \qquad (3)$$

where, $\delta_{ji}$ is the Kronecker delta ($\delta_{ji} = 0$ if $j \neq i$, and $\delta_{ji} = 1$ if $j=i$), and $r_{ji}$ is a random is uniformly distributed between zero and one, ($0 \leq r_m < 1$) and $r$ is a real constant, which is small. In Eq. (3) $r_{ji}$ may be considered as the number of the adults who change their language from their mother tongue ($i$) to another language ($j$) at $t$. (These individuals may have immigrated to a city where the language ($j$) is spoken, or another reason may be decisive.) For simplicity, we take $r_{ji} = r_j \delta_{ji}$, (It is obvious that, $r_{ji}$ may be utilized without the present approximation for computations using super computers.) where $r_j$ is a random number which is uniformly distributed between zero and one so that the multiplication $r\, r_j$ (Eq. (3)) may represent the net (positive) change in the number of the speakers of the language ($j$).

On the other hand, all of the speakers of a language (which emerged newly in terms of fragmentation, for example) may change their language, or they may be colonized, etc. We consider these cases, with a probability $x$. Secondly, when a new language is formed (at $t$), she may belong to the same family as the home language or a new family may be started in the meantime; which means that, the families fragment (by the probability $h_{fam}$). Please note that, the probability of forming a language (belonging to a new family) defines the parameter $h_{fam}$, and the families grow exponentially in number (up to randomness) with time, as the languages do, Eq. (2)).

For the lifetimes we simply subtract the number of the time step when a language is emerged, from the total number of time steps for the ages of the living languages, and we ignore the lifetimes of the extinct languages. The following remark finishes the definition of our model: Whenever a language (family) becomes less then unity in size, we consider her as extinct. (We do not consider punctuation of the languages or of the families with their population, since it is not historically real.)

**3. Application:** We have $m(0)$ ancestor languages, and $f(0)$ ancestor families. The evolution of the size of any (the $j^{th}$) language maybe given as; $L(t+1)_j = \sum_{i=1}^{m(t)} E(t)_{ji} L(t)_i = \sum_{i=1}^{m(t)} (\delta_{ji} + r_{ji}) L(t)_i = \sum_{i=1}^{m(t)} (1 + r_j) \delta_{ji} L(t)_i = L(t)_j + rW(t)/2$, where $W(t)$ is the current world population, i.e.,

$$W(t) = \sum_{i=1}^{m(t)} L(t)_i \quad . \qquad (4)$$

Please note that, the growth rate of a language is proportional to this of the world (up to randomness), since we assume that each human speaks one (mother) language; and, if the languages grow by a random probability between $0$ and $r$ at each time step, then $W(t)$ grows by the same random probability between $0$ and $r$ up to a constant (and, vice versa).

It is obvious that, the introduced rates ($h$, $x$, $h_{fam}$ and $r$) have unit of time, and our time unit is arbitrary. We know that, we have about 7 thousand languages and about 7 billion people living at the present time[3]; when we reach at these numbers (for a given set of parameters) we stop our computation and study the results (for various total number of time steps). Then, we consider the evolution of the number of languages and of the world population, size distribution of the languages and of the families, the distribution of the families over the number of the languages, lifetimes of the languages, etc. We consider also the size distribution of the second languages where we assume that an adult selects a language (else then the mother language) if the second language ($k$) is bigger than the mother language ($i$).

So, the number of people who speak the language (*i*) as the mother tongue and selects of the language (*k*) as the second language at a time *t* is,

$$L'(t)_k \propto L(t)_i (L(t)_k - L(t)_i) \lambda r'_i \quad , \tag{5}$$

where, $L(t)_i \leq L(t)_k$ and the prime denotes the second language. In Eq. (5) $r'_i$ is a random number which is uniformly distributed between zero and one (so; $0 \leq \lambda r'_i < \lambda$ for a given $\lambda$) which is proportional to the percentage (up to randomness) of the population of the language (*i*) the speakers of which select (*k*) as the second language, and $\lambda$ is taken as universal. It is obvious that, $\lambda$ has the unit per capita (person) and as the size difference $(L(t)_k - L(t)_i)$ increases, the language (*k*) becomes more favorite and the related percentage $((L(t)_k - L(t)_i)\lambda r'_i)$ increases.

**4. Results:** Within our computations, we utilize several parameters for *h*, *x*, and *r*, where we keep *s*=0.5, *f(0)*=50 and $h_{fam}$ =0.02 for several number of time steps (200, or less) in all. Please note that, as *x* increases *h* should also increase, for the given number of the languages and world population for the present time; since, increasing *x* *x* (=0; 0.05; 0.1; etc.) decreases the number of the languages. Secondly, we do not vary $h_{fam}$ since, the number of present families is relatively small (about 130 [3], i.e. about 60 times smaller than the number of the present languages) and we predict that, the results are not sensitive to $h_{fam}$ for 200 time steps, or less.

We tried several cases (*m(0)*=50 to 1000, *f(0)*=50, *W(0)*= 250,000 to 5 million) for the initial world, and we predict that the initial parameters are not decisive on the results, where several sets of the parameters for the evolution may be selected (not shown) to mimic the empirical data [3]; i.e., a wide range of initial parameters and a wide range of evolution parameters may give similar results. For example with *m(0)*=50 and *W(0)*=250,000; we have 7740 present languages for *h*=0.027, *x*=0.005, and *r*=0.15 (with the exponents 0.015, and 0.027 for the number of languages and world population, respectively) after 150 time steps. In another example we have *m(0)*=1000 and *W(0)*=5 million and we have *m(t)*= 7734 and *W(t)*= (about) 5 billion, for the present time (the exponents are 0.02, and 0.04 for the number of languages and world population, respectively), where the parameters for the evolution are: *h*=0.012, *x*=0.05, and *r*=0.31 for 45 time steps. So we make the following comment: Several different initial conditions might have taken place in the world and the parameters (*h*, *x*, *r*) for the evolution might be decisive for the survival of the languages.

Figure 1 is the probability distribution function for the languages over size, where the parameters are *m(0)*=50, *W(0)*=250,000, *h*=0.12, *x*=0.05, *r*=0.31 for 110 time steps (others are the same as before). Please note that, Fig. 1 is similar to the empirical data [3] and many more similar ones are obtained (not shown) with different parameters, which diverse over wide range.

Figure 2 is the language families versus the number of the languages (and size, in the inset); where, the axes are logarithmic and the families are placed on the horizontal axes with the rank in decreasing order (i.e., the biggest family (in number in Fig. 2, and in size in the inset) has the rank 1, the next biggest family has the rank 2, etc. The inset is the size distribution of the families. Please note that, Fig. 2 and the inset (for the languages and families) are similar to empirical plots [3], and the present results for the families are better than the ones in [4]; where the number of families were kept as constant (with no fragmentation) but here we have fragmentation for the families with the rate $h_{fam}$=0.02 with *f(0)*=50 (other parameters are the same as in Fig. 1). Now the family distribution over the number of the languages follows power law -2, as the arrow (dashed) in Fig. 2 indicates and the result is in excellent agreement with empirical data [3].

The lifetimes of the living languages (ages) are given in Figure 3 (crucial parameters are given in the related caption for both of the plots), where we have decreasing exponential distribution as in [4]. Please note that, for big $x$ (thin plot, in circles) we have no ancestor language living at the present time.

Figure 4 depicts the size distribution for the second languages, under the assumption made for Eq. (5), where the horizontal axis has an arbitrary unit for the speakers (say, percentage). Please note that, about 99 percent of the people speaking second language is utilizing 100 big languages as the second language in our results (Fig. (4), right hand); where, the parameters are the same as in Fig. 1.

**5.   Discussions and conclusion:**   In Eq. (3) the (random) term $r_{ji}$ may be considered as the connection between human speaking the languages ($j$) and ($i$). In other words, the term $r_{ji}$ with $-r < rr_{ji} < r$ may be utilized for the adults who (immigrate and) change their language from ($j$) to ($i$) for $0 < r_{ji}$ and vice versa for $r_{ji} < 0$, with the probability $r$ in all; which may be taken into account for more precise computations. And, the mentioned movement (mixture) of the people from one language to the other may be one of the possible reasons for the evolution of the languages (linguistically) since humans carry many words and rules of the previous language in mind which may result in several misunderstanding and misuses of several issues of the new language, etc. In terms of such minor changes the languages may become richer and evolve linguistically. Now, the number of the old languages may be multiplied with the number of immigrants and added to the current result of a similar multiplication for the language under consideration and by dividing the result with the total number of the current speakers of the given language, we may assign a new number for the given language. And, if this assigned number changes by a certain amount (say, one), then we may consider a total change in the language, which may also be taken into account for more precise computations.

On the other hand, it is obvious that, $L(n)=E(n)L(0)$, with $E(n)= E(t=n)E(t=n-1)…E(t=1)$ and $L_i(0)=pr_i$; where $r_i$ is a random real number with $0 \leq pr_i < p$, and $p$ is a real constant standing for the maximum number of the speakers of the ancestor languages, and $L(0)$ is a ($1 \times m(0)$) random matrix. The matrix $E(n)$ is rectangular and, its size increases exponentially in time due to the exponential increase in $m(t=n)$ with time, and the period of time for the computation increases exponentially, where organization of $E(n)$ into sub matrices in terms of blocks and utilizing these block matrices separately for the evolution of the languages all of which belong to the same family (represented by the given sub matrix) may be useful for saving the computer time. It is obvious that the present approach in matrix format may be utilized for the biological evolution of the species, and also for the cities, etc., if good computer facilities are available or block matrices are utilized within the software.

As a conclusion, it may be stated that, we have many random parameters for the initial world, and many random parameters for the evolution; yet, the results come out with several regularities, which mimic various empirical data [3] fairly good. We claim that, the present regularities are formed out of randomness by the two opposite processes; random multiplicative noise [1] and random fragmentation [2]. Furthermore, we suggest that, several other processes may be shaping several other regularities out of randomness in the same and in some other fields of science.

**Acknowledgement**
The author is thankful to Dietrich Stauffer for his critical reading and corrections, and valued discussions.

FIGURES

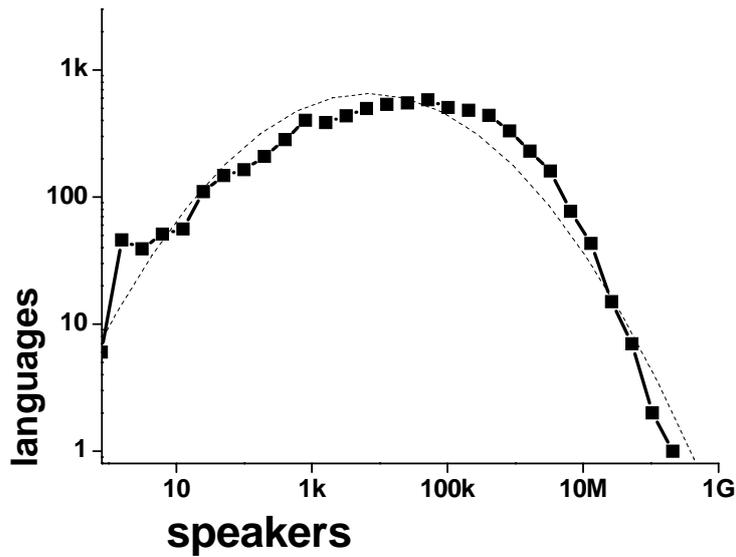

**Figure 1**  Size distribution of the present languages. The dashed line (for a parabolic fit) indicates that the present languages depict slightly asymmetric log-normal distribution. (For the parameters, please see Sect. 4.)

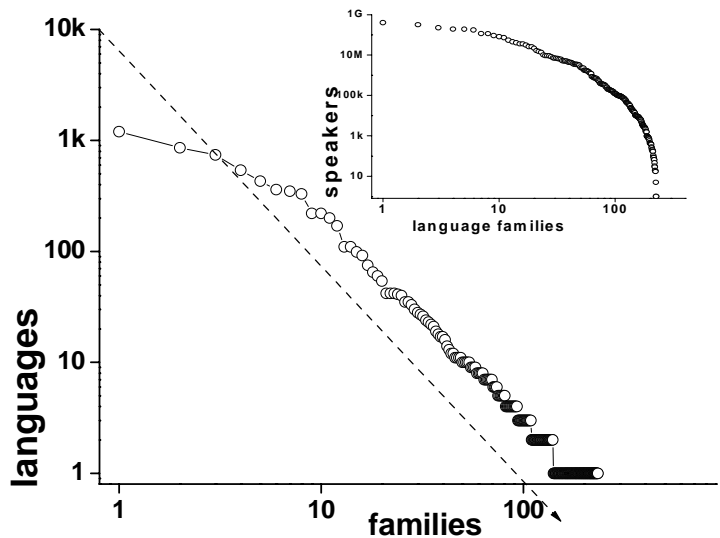

**Figure 2**  The language families (in rank order along the horizontal logarithmic axis) versus the number of the members (along the vertical logarithmic axis) at the present time, where we have about 140 families with 7450 languages. Initially we have 50 languages and 50 of families. The inset is the size distribution of the present families. (For the other parameters, please see Sect. 4.)

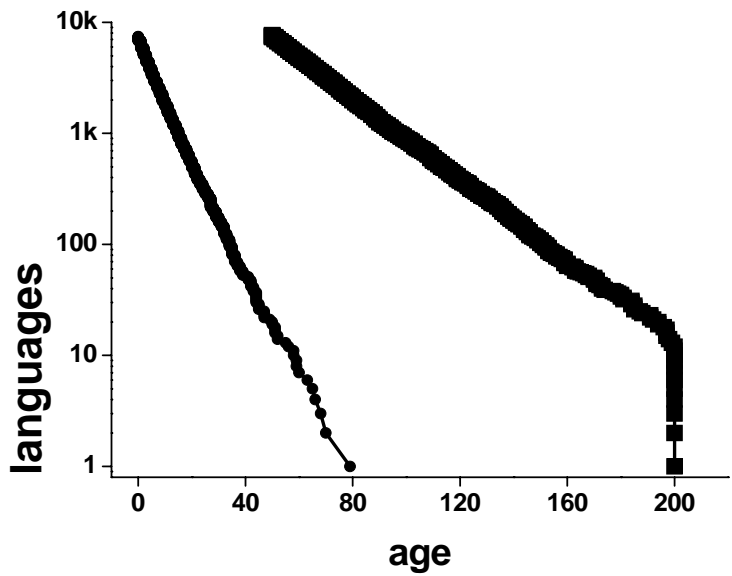

**Figure 3**  Lifetimes (ages) of the present languages. Thin (cirle) plot is for *x*=0.05 (with, *m(0)*=50, *W(0)*=250,000, *h*=0.12 and *r*=0.31 in 110 time steps, for the other parameters please see Sect. 4); and thick (square) plot is for *x*=0.005 (with *h*=0.027, *r*=0.15 in 200 time steps, other parameters are the same as in thin plot.)

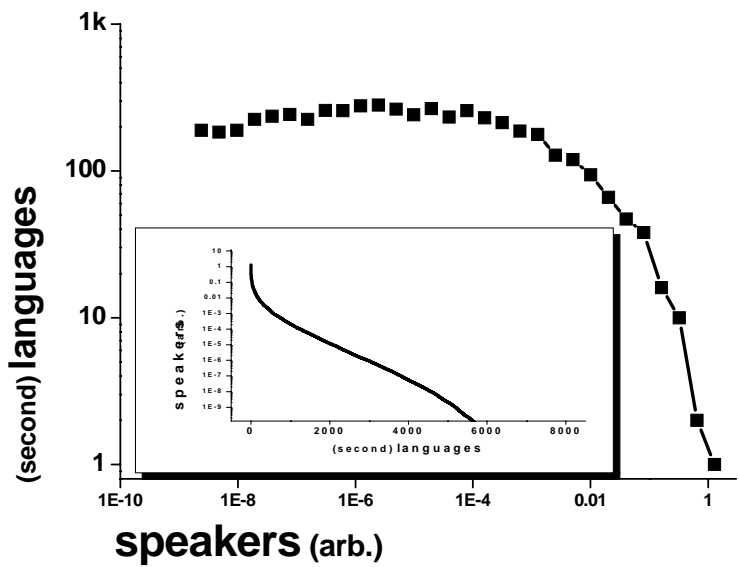

**Figure 4**  The probability distribution function for the second languages over (arbitrary) size, where bigger languages (than the mother languages) are favored (Eq. 5) with $\lambda$=0.01. The inset is the distribution of the second languages over (arbitrary) size, where the languages are in rank order and the vertical axis is logarithmic. (For the other parameters, please see Sect. 4.)